\documentclass[aps,twocolumn,showpacs]{revtex4}
\usepackage{graphicx}
\usepackage{amsfonts}
\usepackage{amssymb}
\usepackage{amsbsy}
\usepackage{amsmath}
\usepackage{mathrsfs}
\usepackage{latexsym}
\usepackage{natbib}
\usepackage{bm}
\usepackage{color}
\usepackage{braket}
\usepackage{slashed}
\usepackage[normalem]{ulem}

\def\k{\mathrm{k}}

\def\gtsigma{\tilde{g}_{\sigma}}

\def\gtomega{\tilde{g}_{\omega}}

\begin{document}

\author{Siddhartha Bandyopadhyay}
\email{sb19ms143@iiserkol.ac.in}

\author{Golam Mortuza Hossain}
\email{ghossain@iiserkol.ac.in}

\affiliation{ Department of Physical Sciences, 
Indian Institute of Science Education and Research Kolkata,
Mohanpur - 741 246, WB, India }
 

%\date{\today}

\title{Probing the equation of state of neutron stars using neutrino
oscillations}

\begin{abstract}
We study the phenomena of neutrino oscillations and flavour mixing by
incorporating the gravitational effects through the Dirac equation in curved
spacetime inside a spherically symmetric star. We show that the flavour
transition probabilities of the neutrinos depend on the interior spacetime
metric as they propagate out of the star. As a consequence, we show that one 
could distinguish between different possible equation of states of nuclear 
matter even for an isolated neutron star if one could determine the flavour 
composition of emitted neutrinos near the stellar surface.
\end{abstract}

\maketitle

\section{Introduction}

The phenomena of neutrino oscillations wherein neutrinos of one flavour may
convert into those of other flavours while propagating through the vacuum or a
matter medium, has been observed in experiments such as MINOS, SNO \cite{MINOS,
SNO} and is still being routinely observed in experiments such as T2K, KamLAND,
Super-Kamiokande, NOvA and IceCube \cite{T2K, KamLAND, SKev, Wester_2024, NOvA,
Abbasi_2023}. In our current understanding, the phenomena of neutrino oscillations
requires neutrinos to be massive particles and it indicates possible existence
of new physics beyond that of the Standard Model of particle physics.

For theoretical modelling of neutrino oscillations, one usually employs the
Minkowski spacetime. However, there are circumstances wherein neutrinos may
propagate through the regions of spacetime where gravitational field is 
extremely strong such as in the interior of neutron stars. In fact, the neutron
stars are known to be a major source of astrophysical neutrinos, produced
through the $\beta$-equilibrium processes in the nuclear matter. For such
situations, one needs a framework to take into account the effects of
curved spacetime when calculating flavour transition probabilities. In the
existing literature, indeed, the effect of gravitational fields on neutrino
oscillations and flavour mixing have been considered by using either quantum
mechanical methods or by using the Dirac equation in curved spacetime with varied
degrees of rigour \cite{Ahluwalia_1996, ahluwalia1996interpretation,
Grossman_1997, P_riz_1996, Heuristic_treatment, Tanmoy, maiwa2004,
G_D__2011, Zhang:2016deq, mandal2021neutrino, Capolupo:2023wri}.

In this article, our aim is two fold. Firstly, we compute the flavour
transition probabilities of neutrinos in a spherically symmetric stellar
spacetime by employing the methods of quantum field theory in curved spacetime.
It leads to the clarification regarding the notion of conserved energy
for propagating neutrinos in a curved spacetime. Subsequently, we consider
the matter effects on the transition probabilities, referred to as the
Mikheyev-Smirnov-Wolfenstein (MSW) effect \cite{Mikheev:1987jp, Smirnov_2005,
Smirnov2003}, in the given curved spacetime.

Secondly, we show that the phenomena of neutrino oscillations could be used to
distinguish between different possible equation of states (EOS) of nuclear
matter contained in the interior of a neutron star if we could determine the
flavour composition of emitted neutrinos, particularly the presence of
non-electron neutrinos, near the stellar surface. It may be mentioned that in
$\beta$-equilibrium, the constituent nucleons go through the processes of
$\beta$-decay and inverse $\beta$-decay, known as the Urca processes. These 
processes produce neutrinos or anti-neutrinos of electron flavour only. 
The suggested method here could be used even for an isolated neutron star
where constraining its nuclear matter EOS through the tidal deformation method
is not available.

\section{Spacetime Metric}\label{sec:InteriorSpacetime}

In order to describe the astrophysical neutron stars which are known to be
rotating bodies, ideally one should consider the spacetime to be described by an
axially symmetric metric such as those studied in \cite{hartle1967slowly,
hartle1968slowly}. The effect of inertial frame-dragging arising due to the
rotation of compact bodies such as the neutron stars, can lead to important
consequences \cite{hossain2024origin}. However, the effect of inertial
frame-dragging  on the equation of state has been shown to be extremely small
\cite{hossain2021equation}. Therefore, to describe the interior and exterior
spacetime of a neutron star, for simplicity here we consider a spherically
symmetric metric that can be represented, in the \emph{natural units} $c = \hbar
=1$, by an invariant line element as
\begin{equation}\label{SphericallySymmetricMetric}
ds^2 = -e^{2\Phi}dt^2 + e^{2\nu}dr^2 + r^2[d\tilde{\theta}^2
+\sin^2\tilde{\theta} d\varphi^2 ] ~.
\end{equation}
Here the metric functions $\Phi = \Phi(r)$ and $\nu = \nu(r)$ are functions of
the radial coordinate $r$ only.

In order to express the Einstein equation, we model the stellar matter to be
described by a perfect fluid whose stress-energy tensor is given by
\begin{equation}\label{Fluid tensor}
T_{\mu\nu} = (\rho+P)u_{\mu} u_{\nu} + P g_{\mu\nu}  ~,
\end{equation}
where $u^\mu$ is the 4-velocity of the stellar fluid satisfying $u^\mu
u_\mu=-1$, $\rho$ is the energy density and $P$ is the pressure of the fluid.
If we introduce a mass function $\mathcal{M}=\mathcal{M}(r)$ in terms of the
metric function $\nu(r)$ so that  $e^{-2\nu} = (1- 2G\mathcal{M}/r)$ then the
Einstein equation corresponding to the metric (\ref{SphericallySymmetricMetric})
implies $(d\mathcal{M}/dr)=4\pi r^2\rho$. Furthermore, the equations for the
metric function $\Phi$ and the pressure $P$ are given by
\begin{equation} \label{TOVEq}
 \frac{d\Phi}{dr} = \frac{G(\mathcal{M}+4\pi r^3 P)}{r(r-2G\mathcal{M})}
~,~ \frac{dP}{dr} = -(\rho+P)\frac{d\Phi}{dr}  ~.
\end{equation}
The equations (\ref{TOVEq}) are referred to as the Tolman-Oppenheimer-Volkoff
(TOV) equations \cite{Tolmann, TOV1939PhRv...55..374O}. The exterior vacuum
metric can be solved exactly as
\begin{equation}\label{ExteriorMetric}
e^{2\Phi(r)} = e^{-2\nu(r)} = 1- \frac{2GM}{r}  ~,
\end{equation}
where $M$ represents the total mass of the star and $G$ is Newton's constant of
gravitation. Consequently, the interior metric solutions of the TOV equations
(\ref{TOVEq}) are subject to the boundary condition $e^{2\Phi(R)} =
(1-2GM/R)$ where $M = \mathcal{M}(R)$ with $R$ being the radius of the star.

\subsection{Conserved energy of a particle}

In the study of neutrino oscillations in flat spacetime, one needs to 
characterize the neutrino states having definite energy eigenvalues. Therefore,
in the curved spacetime, it is essential to define an invariant notion of energy
in order to describe the trajectory of a neutrino which is considered to be a
massive particle. In general, if a spacetime metric $g_{\mu\nu}$ admits a 
Killing vector field $t^{\mu}$ then for a particle of mass $m$, moving along a 
geodesic with the 4-momentum $p^{\mu}= m u^{\mu}$, one can define a conserved 
quantity $\varepsilon = - t^{\mu} p_{\mu}$ which satisfies
\begin{equation}\label{EnergyConservationEq}
\frac{d\varepsilon}{d\tau} = - m \left[ u^{\alpha} u^{\beta} \nabla_{\alpha}
t_{\beta} +  t^{\beta} u^{\alpha} \nabla_{\alpha} u_{\beta} \right] = 0 ~,
\end{equation}
provided $t^{\mu}$ satisfies the Killing equation $\nabla_{(\alpha}t_{\beta)} =
0$
and 4-velocity $u^{\mu}$ satisfies the geodesic equation $u^{\alpha}
\nabla_{\alpha} u_{\beta} = 0$. Here $\tau$ refers to the associated proper
time.

\subsubsection{In the interior of a star}

The spherically symmetric metric (\ref{SphericallySymmetricMetric}) admits a 
timelike Killing vector of the form $t^{\mu} = (1,0,0,0)$. Consequently, the 
associated conserved quantity $\varepsilon= - p_0$ can be viewed as the energy 
of the particle by an asymptotic observer. We note that for a \emph{radially} 
moving particle through the interior spacetime of a star, the relation 
$g_{\mu\nu} p^{\mu} p^{\nu} = - m^2$ implies
\begin{equation}
\label{EnergyMomentumInterior}
\varepsilon = e^{\Phi} \sqrt{(e^{\nu} p^{r})^2 + m^2}  ~,
\end{equation}
where $p^r = m u^r = (mv)u^0$ with radial 3-velocity being $v=(dr/dt)$. Here
proper time $\tau$ is defined as $d\tau^2 = -ds^2 = dt^2(e^{2\Phi} - e^{2\nu}
v^2)$ along a timelike trajectory.

\subsubsection{In the exterior of a star}

In the exterior regions of a star the conserved energy expression
(\ref{EnergyMomentumInterior}) reduces to
\begin{equation}\label{EnergyNewtonianLimit}
\varepsilon \simeq m  + \frac{1}{2} m v^2 - \frac{GMm}{r} ~,
\end{equation}
when the gravity is weak \emph{i.e.} $(GM/r) \ll 1$ and the particle is 
moving with a non-relativistic speed \emph{i.e.} $v^2 \ll 1$. The equation
(\ref{EnergyNewtonianLimit}) makes it clear that for an asymptotic observer the
quantity $\varepsilon$ indeed represents the Newtonian total energy, including 
rest energy ($mc^2$ with $c=1$), of a particle moving radially in the
gravitational field of a star having mass $M$. Further, we note that in the flat
spacetime limit ($\Phi\to 0$, $\nu\to 0$) the expression of the conserved energy
(\ref{EnergyMomentumInterior}) reduces to its Minkowskian expression for a
relativistic particle given by $\varepsilon = \sqrt{(p^{r})^2 + m^2}$.

\section{Neutrino Oscillations}\label{sec:NeutrinoOscillation}

In the Standard Model of particle physics, there are \emph{three} flavours of
neutrinos namely electron neutrino, muon neutrino and tau neutrino. We denote
these \emph{flavour} states as $\nu_e$, $\nu_{\mu}$ and $\nu_{\tau}$
respectively. Furthermore, in the Standard Model these neutrinos are considered
to be massless particles. However, in order to explain the current
observations, one needs to consider these neutrinos to be massive particles.

\subsection{Mixing angle}

The flavour states of the neutrinos are considered to be linear superpositions
of the energy (mass) eigenstates of neutrinos, say $\psi_j$, having masses
$m_j$ for $j=1,2,3$. In other words, the flavour states can be expressed as
$\nu_{\alpha} = \sum_{j} {U_{\alpha j}} \psi_j$ for $\alpha=e,\mu,\tau$. The
unitary matrix $U_{\alpha j}$ is known as the Pontecorvo-Maki-Nakagawa-Sakata
(PMNS) matrix.
The basic idea behind the neutrino oscillations, nevertheless, can be understood
by considering the simpler case of 2-flavour neutrino oscillations
\cite{GRIBOV1969493, 10.1143/PTP.28.870}. Here we consider two flavour states
namely $\nu_e$ and $\nu_a$, where $a$ denotes non-electron flavour neutrinos
which could be either $\mu$, $\tau$ neutrinos or even their linear
superposition. The flavour states then can be expressed as a superposition of
the mass eigenstates by using the so-called \emph{mixing angle} $\theta$ as
\begin{equation}\label{PMNS}
    \begin{pmatrix}
        \nu_e \\ \nu_a
    \end{pmatrix}=
        \begin{pmatrix}
            \cos\theta  & \sin\theta \\
            -\sin\theta  & \cos\theta
        \end{pmatrix}
    \begin{pmatrix}
        \psi_1 \\ \psi_2
    \end{pmatrix} ~.
\end{equation}
In vacuum, the angle $\theta$ is referred to as the vacuum mixing angle whereas
in the presence of matter the angle $\theta$ is referred to as the matter mixing
angle and is usually denoted as $\theta_m$.

The energy eigenstates of the neutrinos follow the time-evolution equations
$i\partial_t \psi_1 = \varepsilon_1\psi_1$ and $i\partial_t \psi_2 =
\varepsilon_2\psi_2$ where $\varepsilon_1$ and $\varepsilon_2$ are the energy
eigenvalues. Using the mixing equation (\ref{PMNS}), we may express the
time-evolution equation for the flavour states as
\begin{equation}\label{FlavourEvolutionEqFlat}
i \partial_t
\begin{pmatrix}
  \nu_e \\ \nu_a
\end{pmatrix}
= \left[
 \varepsilon \mathbb{I} - \tfrac{\Delta\varepsilon}{2}
\begin{pmatrix}
  -\cos2\theta  & \sin2\theta \\
 \sin2\theta  & \cos2\theta
\end{pmatrix} \right]
\begin{pmatrix}
  \nu_e \\ \nu_a
\end{pmatrix}   ~,
\end{equation}
where $\varepsilon = \tfrac{1}{2}(\varepsilon_1 + \varepsilon_2)$ and 
$\Delta\varepsilon = (\varepsilon_1 - \varepsilon_2)$. We note that the average 
energy term $\varepsilon$ leads to an overall phase of the flavour states 
whereas the energy gap $\Delta\varepsilon$ leads to an oscillation between
the flavour states.

\subsection{Phase difference due to the propagation}

During the propagation through the spacetime, the energy eigenstates evolve as 
$\psi_j(t) = \psi_j(0) e^{-i\phi_j}$ where $\phi_j = \int dt\varepsilon_j$ for 
$j=1, 2$. It results in a net phase difference between these energy eigenstates 
leading to a transformation between the neutrino flavour states. For 
definiteness, let us consider a beam of neutrinos which is produced at a source 
located at $r=0$ and then it propagates radially outward over a baseline say 
$L$. The resultant phase difference can be expressed as
\begin{equation}\label{DeltaPhaseDef}
\Delta \phi_{osc} = \phi_1 - \phi_2 = \int_0^L dr
\Big[\Big(\frac{dt}{dr}\Big)_1 \varepsilon_1
- \Big(\frac{dt}{dr}\Big)_2 \varepsilon_2 \Big]  ~.
\end{equation}
Along a radial geodesic, we can express $(dt/dr) = (e^{-2\Phi}
\varepsilon/p^{r})$ where $\varepsilon$ is conserved but radial momentum   $p^r
= e^{-\nu} \sqrt{(e^{-\Phi} \varepsilon)^2 - m^2}$ is not a conserved quantity.
If the mass is very small compared to the energy \emph{i.e.} $(m/\varepsilon)^2
\ll 1$ then we can approximate
\begin{equation} \label{GeodesicDTDR}
\left(\frac{dt}{dr}\right) = ~e^{\nu - \Phi} \left[1 +
\frac{m^2 e^{2\Phi}}{2 \varepsilon^2}
+ \mathcal{O} \left( \frac{m^4}{\varepsilon^4} \right) \right]  ~.~
\end{equation}
By using the approximation (\ref{GeodesicDTDR}), we can express the phase
difference between two energy eigenstates (\ref{DeltaPhaseDef}) in a spherically
symmetric spacetime as
\begin{equation}\label{DeltaPhaseCurvedFull}
\Delta \phi_{osc} \simeq
\Big(\frac{m_1^2}{2\varepsilon_1}-\frac{m_2^2}{2\varepsilon_2}\Big)
\int_0^L dr~e^{\Phi + \nu} + \Delta\varepsilon \int_0^L dr~e^{\nu - \Phi} ~.
\end{equation}
The phase difference (\ref{DeltaPhaseCurvedFull}) during propagation is
controlled by $4$ \emph{invariant} parameters namely $\varepsilon_1$, 
$\varepsilon_2$, $m_1$ and $m_2$ in general. However, only certain combinations 
of these $4$ parameters appear in the phase difference.

If the energy gap between the two energy eigenstates is small \emph{i.e.}
$(\Delta\varepsilon/\varepsilon) \ll 1$ but squared mass gap
$\Delta m^2 = (m_1^2 - m_2^2)$ is not very small \emph{i.e.} $\Delta m^2 \sim
(m_1^2 + m_2^2)$ then we can approximate the phase difference
(\ref{DeltaPhaseCurvedFull}) as
\begin{equation}\label{DeltaPhaseCurvedA1}
\Delta \phi_{osc} \simeq \frac{\Delta m^2}{2\varepsilon} \int_0^L dr~e^{\Phi +
\nu} + \Delta\varepsilon \int_0^L dr~e^{\nu - \Phi} ~.
\end{equation}
The phase difference (\ref{DeltaPhaseCurvedA1}) has two different types of 
gravitational contributions. Firstly, even if neutrinos were massless there 
would be a gravitational correction to the phase difference as long as 
the energy gap $\Delta\varepsilon$ is non-zero. Secondly, the phase difference 
with a gravitational correction would arise even if the energy gap
$\Delta\varepsilon = 0$ but there is a non-zero squared mass gap $\Delta m^2$.

\subsubsection{The flat spacetime limit}

\begin{figure}
\includegraphics[height=5.9cm, width=8cm]{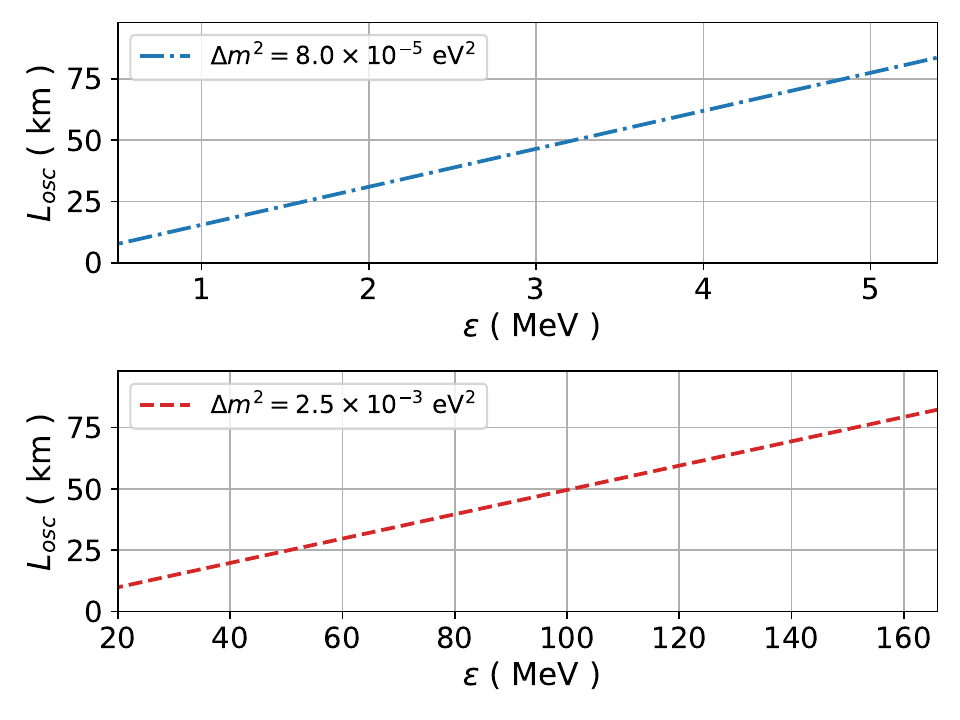}
\caption{The vacuum oscillation length of neutrinos for different mass squared
difference $\Delta m^2$  and energy $\varepsilon$.}
\label{fig:VacuumOscLength}
\end{figure}

In the flat spacetime approach, the flavour states are considered to be the
superposition of the energy eigenstates often having same radial momentum
$p_1^r = p_2^r$ which implies $\Delta\varepsilon \simeq (\Delta
m^2/2\varepsilon)$\cite{GIUNTI_2001}. Consequently, in the flat spacetime limit
\emph{i.e.} $\Phi\to 0, \nu\to 0$, phase difference (\ref{DeltaPhaseCurvedA1})
reduces to
\begin{equation}
\label{DeltaPhaseFlat}
\Delta\phi_{osc}
\simeq \Big(\frac{\Delta m^2}{2\varepsilon} + \Delta\varepsilon\Big)L
\simeq \frac{\Delta m^2 L}{\varepsilon}
= 2 \pi \Big(\frac{L}{L_{osc}}\Big) ~,~
\end{equation}
where $L_{osc} = (2\pi\varepsilon/\Delta m^2)$ is known as the vacuum 
oscillation length of neutrinos. The vacuum oscillation lengths for different 
combinations of the parameters $\Delta m^2$ and $\varepsilon$ are shown in the
FIG. \ref{fig:VacuumOscLength}.

\subsubsection{A missing factor of $2$}

We note that the flat spacetime expression of $\Delta\phi_{osc}$ 
(\ref{DeltaPhaseFlat}) in terms $(\Delta m^2 L/\varepsilon)$  differs by a 
factor of $2$ compared to the analogous expression often stated in many 
literatures \cite{G_D__2011, Bilenky} (however see \cite{Tanmoy}). This 
difference essentially arises because in those literature one assumes, rather 
erroneously, that a neutrino follows a null trajectory \emph{i.e.} $(dt/dr) = 
1$. Neutrino being a massive particle, albeit with a very small mass, it 
follows a timelike trajectory given by (\ref{GeodesicDTDR}). This aspect of the
neutrino trajectory leads to an additional contribution involving $(\Delta 
m^2/2\varepsilon)$ to the phase difference (\ref{DeltaPhaseCurvedA1}, 
\ref{DeltaPhaseFlat}).

\section{Fermions in curved spacetime}

In the previous section, we have considered the neutrinos as relativistic, 
massive particles that are propagating along the geodesics in a curved 
spacetime. However, the neutrinos are fermions and are described by the spinor 
field in the Standard Model. Therefore, in order to describe the propagation of 
neutrinos in a curved spacetime, it is imperative to employ a generally 
covariant action for the fermions.

\subsection{Dirac equation}

We consider the neutrino to be a Dirac spinor here. In the Fock-Weyl
formulation, the dynamics of a free fermion in a curved spacetime is governed by
the action \cite{Weyl:1929fm,Fock}
\begin{equation}\label{DiracActionCurved}
S_{\psi} = - \int d^4x \sqrt{-g} \bar{\psi}[i\gamma^a {e^{\mu}}_a
\mathcal{D}_{\mu} + m] \psi ~,
\end{equation}
where $\bar{\psi}=\psi^\dagger \gamma^0$ is known as the Dirac adjoint and $m$
being the mass of the fermion. Here ${e^{\mu}}_a$ are the components of the
tetrads which form a basis for the local tangent space and relates it to the
global coordinates such that $g_{\mu\nu} {e^{\mu}}_a {e^{\nu}}_b = \eta_{ab}$
where $\eta_{ab} = diag(-1,1,1,1)$. The spin covariant derivative is given by
$\mathcal{D}_\mu \psi = \partial_\mu \psi + \Gamma_\mu \psi$ and after imposing
the compatibility conditions for tetrads, i.e. $\mathcal{D}_\mu
{e^{\nu}}_a = 0 = \mathcal{D}_\mu e_{\nu}^{\;\;a}$ we get the spin connection
explicitly as
\begin{equation} \label{SpinConnectionDef}
\Gamma_\mu = -\frac{1}{8} \eta_{ac} {e_{\nu}}^c (\partial_\mu
{e^{\nu}}_b + \Gamma^\nu_{\mu\sigma} {e^{\sigma}}_b ) [\gamma^a,\gamma^b] ~,
\end{equation}
where $\Gamma^\nu_{\mu \sigma}$ are the Christoffel connections and $\gamma^a$
are the standard Dirac matrices in Minkowski spacetime satisfying the Clifford
algebra $\{\gamma^a,\gamma^b\}=-2\eta^{ab}\mathbb{I}$ such that
$(\gamma^0)^2 = \mathbb{I}$ and $(\gamma^k)^2 = -\mathbb{I}$ for spatial
indices $k= 1,2,3$. The conserved 4-current corresponding to the
action (\ref{DiracActionCurved}) is $j^\mu=\Bar{\psi}\gamma^a e^\mu_{\;\;a}\psi$
which satisfies $\mathcal{D}_\mu j^{\mu}=\nabla_\mu j^\mu=0$.

The generally covariant field equation for the spinor field \emph{i.e.} 
$[i\gamma^a {e^{\mu}}_a \mathcal{D}_{\mu} + m] \psi = 0$, can also be expressed 
as
\begin{equation}
\label{DiracEqTimeEvolutionForm}
i\partial_t \psi = - \left[
{e_t}^0 \gamma^0 \left( i\gamma^a {e^{j}}_{a} \mathcal{D}_j + m \right) +
i\Gamma_t \right] \psi  \equiv  \tilde{H}^0 \psi ~,
\end{equation}
where $j$ is summed over the spatial indices. The form of the field equation 
(\ref{DiracEqTimeEvolutionForm}) resembles a quantum mechanical evolution 
equation with $\tilde{H}^0$ being the Hamiltonian operator. However, in a curved 
spacetime the operator $\tilde{H}^0$ may not hermitian in general
\cite{PARKER1}.

\subsection{Dirac equation inside a spherical star}

In order to reduce the Dirac equation (\ref{DiracEqTimeEvolutionForm}) into a 
form which is  directly applicable for a spherical star, we need to choose the 
appropriate tetrads and then compute the corresponding spin connections.

\subsubsection{Tetrads and spin connections}

By making a diagonal ansatz, the non-vanishing tetrad components
corresponding to the metric (\ref{SphericallySymmetricMetric}) can be
expressed as
\begin{equation}\label{TetradSS}
{e^t}_0 = e^{-\Phi}  ~,~
{e^r}_1 =  e^{-\nu} ~,~
{e^{\tilde{\theta}}}_2 =  \frac{1}{r}  ~,~
{e^{\varphi}}_3 =  \frac{1}{r\sin\tilde{\theta}}  ~.
\end{equation}
Similarly, the non-zero components of the inverse tetrad are
\begin{equation}\label{InverseTetradSS}
{e_t}^0 = e^{\Phi}  ~,~
{e_r}^1 =  e^{\nu} ~,~
{e_{\tilde{\theta}}}^2 = r ~,~
{e_{\varphi}}^3 =  r\sin\tilde{\theta}  ~.
\end{equation}
Consequently, the spin-connections (\ref{SpinConnectionDef}) associated with the
spherically symmetric metric (\ref{SphericallySymmetricMetric}) become
\begin{equation}\label{SpinConnectionTRT}
 \Gamma_{t} = \frac{\Phi' e^{\Phi-\nu}}{4}[\gamma^{0},\gamma^{1}] ~,~
 \Gamma_{r} = 0 ~,~
 \Gamma_{\tilde{\theta}} = \frac{e^{-\nu}}{4}[\gamma^{1},\gamma^{2}] ~,
\end{equation}
and
\begin{equation}\label{SpinConnectionP}
 \Gamma_{\varphi} = \frac{\sin\tilde\theta e^{-\nu}}{4}[\gamma^{1},\gamma^{3}]
 + \frac{\cos\tilde\theta}{4}[\gamma^{2},\gamma^{3}]  ~.
\end{equation}
Here we have used the representation of Dirac matrices $\gamma^a$ given by
\begin{equation}\label{DiracMatrices}
\gamma^{0} = \begin{bmatrix}
\mathbb{I}_{2} & 0\\
0 & - \mathbb{I}_{2}
\end{bmatrix}, \ \gamma^{k} = \begin{bmatrix}
0 & \sigma^{k}\\
-\sigma^{k} & 0
\end{bmatrix} ~,
\end{equation}
where $\sigma^{k}$ for $k=1,2,3$ are the Pauli matrices.

\subsubsection{Energy eigenvalues}

In order to arrive at the particle description of neutrinos from the field
equation (\ref{DiracEqTimeEvolutionForm}), we need to ensure the associated
Hamiltonian is hermitian. For a stationary metric, such as the spherically
symmetric metric being considered here, the Hamiltonian can be made
hermitian \cite{PARKER2,6} by using the inner product defined in the curved
spacetime. Consequently, one can associate the quantum-mechanical probability
interpretation with the spinor $\psi$.

We note that the spin connections (\ref{SpinConnectionTRT}, 
\ref{SpinConnectionP}) satisfy $\Gamma_{t}^{\dagger} = \Gamma_{t}$, 
$\Gamma_{\tilde\theta}^{\dagger} = -\Gamma_{\tilde\theta}$ and 
$\Gamma_{\varphi}^{\dagger} = -\Gamma_{\varphi}$. Further, $\Gamma_{\varphi}$ 
and $\Gamma_{\tilde\theta}$ are non-vanishing even in the Minkowski limit due to 
the usage of spherical polar coordinates. Therefore, to ensure hermiticity, we 
may define $H^0 = \tfrac{1}{2}(\tilde{H}^0 + \tilde{{H}^{0}}^{\dagger})$ as the 
hermitian Hamiltonian for describing the propagation of neutrinos. Ideally, 
one should consider a wave-packet for describing propagation of neutrino as a 
particle. For simplicity here we consider the neutrinos to be moving radially 
with the radial wave-vector $\k_r$ such that $i\partial_r \psi_{\k} = \k_r
\psi_{\k}$. The corresponding equation of motion for the $\k^{th}$ mode can then
be expressed as
\begin{equation}
\label{ModeTimeEvolutionEq}
i\partial_t \psi_{\k} = - e^{\Phi} \gamma^0 \left( e^{-\nu} \gamma^1 \k_{r} +
m \right)\psi_{\k} \equiv  H^{0}_{\k} \psi_{\k} ~,
\end{equation}
where $H^{0}_{\k}$ is the Hamiltonian for the $\k^{th}$ mode. By
diagonalising the mode Hamiltonian matrix $H^{0}_{\k}$ we can find the
corresponding energy eigenvalues as
\begin{equation}
\label{ModeEnergyEigenvalue}
\varepsilon_{\k} = \pm e^{\Phi} \sqrt{(e^{-\nu} \k_{r})^2 + m^2}  ~,
\end{equation}
where negative eigenvalues correspond to the anti-particles. We note that if 
we identify the radial wave co-vector $\k_r$ with the radial co-momentum $p_r =
e^{2\nu} p^r$, the energy eigenvalues $\varepsilon_{\k}$ have the same form as 
the conserved energy $\varepsilon$ (\ref{EnergyMomentumInterior}). Besides, 
these modes satisfy the usual quantum mechanical time evolution equation 
(\ref{ModeTimeEvolutionEq}). In other words, these modes follow exactly the same 
properties as the propagating relativistic quantum particles as studied in the 
Section \ref{sec:NeutrinoOscillation}. Henceforth, for brevity of notation we 
shall drop the sub-script $\k$ while describing these modes.

\subsubsection{Transition probability of neutrinos}

Let us consider a beam of electron neutrinos which is produced at a source
located at $r=0$. So the transition probability of an initially
electron neutrino state $\nu_{e|0} = \cos\theta \psi_1(0) + \sin\theta
\psi_2(0)$ transforming into a non-electron neutrino state state $\nu_{a|L} =
-\sin\theta \psi_1(L)  + \cos\theta \psi_2(L) $ after propagating through a
baseline $L$ can be expressed as
\begin{equation}\label{TransitionProbabilityDef}
P(e\rightarrow a)= |\langle \nu_{a|L} | \nu_{e|0} \rangle|^2
= \sin^2 2\theta \sin^2(\Delta\phi_{osc}/2) ~,
\end{equation}
where $\Delta\phi_{osc}$ is the phase difference (\ref{DeltaPhaseDef}) arising
due to the propagation and the angle $\theta$ refers to the vacuum mixing angle
(\ref{PMNS}).

\subsection{MSW Effects}

In the study of solar neutrinos, one includes the effect of matter interaction
by considering the elastic forward scattering (refraction) of neutrinos
\cite{Mikheev:1987jp, Smirnov_2005,Smirnov2003}. This effect is known as the
Mikheyev-Smirnov-Wolfenstein (MSW) effect. In the Glashow-Salam-Weinberg theory
of electro-weak interactions \cite{1967PhRvL..19.1264W}, the neutrinos interact
via 4-fermion current-current interaction terms \cite{WOLFENSTEIN197595} having
both the neutral current and the charged current. However the neutral current 
interaction, arising due to the interaction with the nucleons, is `flavour 
blind' \cite{7} and consequently it leads to an overall phase without impacting 
the neutrino oscillations.

\subsubsection{Charged current interaction}

On the other hand, the charged current interaction affects the neutrino
oscillations and is described by the action
\begin{equation}\label{InteractionAction}
S_{int} = -\int d^4x \sqrt{-g} (\tfrac{G_F}{\sqrt{2}})
[j_{\mu} {\bar{\nu}_e {e^{\mu}}_{a} \gamma^a (1-\gamma^5)
\nu_e}] ~,
\end{equation}
where $G_F$ is the Fermi coupling constant and $j^{\nu} = [{\bar{\psi}_e 
{e^{\nu}}_{b} \gamma^b (1-\gamma^5) \psi_e}]$ is the charged current due to 
the electron field $\psi_e$. In a spherically symmetric, static spacetime, we 
may approximate the background electron current as $j^{\nu} = n_e {e^{\nu}}_0$
where $n_e = n_e(x)$ is the electron number density. Consequently, the
interaction Lagrangian density, defined as $S_{int} \equiv \int d^4x \sqrt{-g}~
\mathcal{L}_{int}$, reduces to
\begin{equation}\label{InteractionLagrangian}
\mathcal{L}_{int} = (\sqrt{2} G_F n_e) [{\bar{\nu}_e \gamma^0 P_L \nu_e}] ~,
\end{equation}
where the projection operator is $P_L = \tfrac{1}{2}(1-\gamma^5)$ and we have
used the fact $g_{\mu\nu} {e^{\mu}}_{a} {e^{\nu}}_{b} = \eta_{ab}$. We note that
$\mathcal{L}_{int}$ does not contain any explicit dependence on the spacetime
metric. However, the metric dependence is contained in the term $\sqrt{-g}$.

The Dirac field equations for the neutrinos having masses $m_1$ and $m_2$ can 
then be expressed as
\begin{equation}
\label{DiracEqTimeEvolutionFormMatter}
i\partial_t \psi_1 = H_1 \psi_1 + H_{12} \psi_2  ~,~
i\partial_t \psi_2 = H_2 \psi_2 + H_{21} \psi_1  ~,~
\end{equation}
where $H_1 = H_1^0 + V \cos^2\theta P_L$, $H_2 = H_2^0 + V \sin^2\theta
P_L$, $H_{12} = H_{21} = \tfrac{V}{2} \sin2\theta P_L$ with $V = (\sqrt{2} G_F 
n_e e^{\Phi})$. Here $H_{1}^0$ and $H_{2}^0$ are the mode Hamiltonians, defined 
in the equation (\ref{ModeTimeEvolutionEq}) for the masses $m_1$ and $m_2$ 
respectively. The presence of the projector $P_L$ in the equation 
(\ref{DiracEqTimeEvolutionFormMatter}) signifies the fact that the electroweak 
sector in the Standard Model is described by $SU(2)_L \times U(1)$ gauge 
symmetry group where only the left-handed neutrino states can interact with the 
background electrons.

\subsubsection{Energy eigenvalues}

Assuming $V$ to be \emph{time-independent}, we can diagonalise the equations
(\ref{DiracEqTimeEvolutionFormMatter}) to obtain the energy eigenvalues in the
presence of matter interaction. For the modes having radial co-momentum $\k_r$,
we can express the interacting energy eigenvalues in terms of the 
non-interacting energy eigenvalues $\varepsilon_1$, $\varepsilon_2$ as
\begin{equation}
\label{MatterEnergyEigvalues}
\varepsilon_{m1} = \varepsilon + \tfrac{1}{2} ( V + \Delta\varepsilon
\mathcal{A})  ~,~
\varepsilon_{m2} = \varepsilon + \tfrac{1}{2} ( V - \Delta\varepsilon
\mathcal{A})  ~,~
\end{equation}
where $\varepsilon = \tfrac{1}{2}(\varepsilon_1 + \varepsilon_2)$ and
$\Delta\varepsilon = (\varepsilon_1 - \varepsilon_2)$. The energy gap in the
presence of matter interaction can be expressed as $\Delta\varepsilon_m \equiv
\varepsilon_{m1} - \varepsilon_{m2} = \mathcal{A} ~ \Delta\varepsilon$.
Here we may define a characteristic number density for the electrons
$n_{\varepsilon} \equiv (\Delta\varepsilon/\sqrt{2}G_F e^{\Phi})$ which 
signifies the strength of the interaction and it allows one to express 
$\mathcal{A}$ as
\begin{equation}
\label{MatterA}
\mathcal{A} = \sqrt{ \left(\cos2\theta - \frac{n_e}{n_{\varepsilon}} \right)^2
+ \sin^22\theta} ~.
\end{equation}
In the flat spacetime limit $\Phi\to 0$, the equations 
(\ref{MatterEnergyEigvalues}, \ref{MatterA}), become identical to those derived
by Wolfenstein \cite{5}. Further, if we turn off the matter interaction by
setting $V\to0$, then $\mathcal{A}\to 1$ and the energy eigenvalues
$\varepsilon_{m1}$ and $\varepsilon_{m2}$ reduce to the non-interacting
eigenvalues $\varepsilon_{1}$ and $\varepsilon_{2}$ as expected.

\subsubsection{Matter mixing angle}

Using the equation (\ref{PMNS}), we can express the time-evolution  equation
(\ref{DiracEqTimeEvolutionFormMatter}) for the neutrino flavour states in the 
presence of matter interaction with a constant potential $V$, as
\begin{equation}\label{FlavourEvolutionEqMatter}
i \partial_t
\begin{pmatrix}
  \nu_e \\ \nu_a
\end{pmatrix}
= \left[\varepsilon_m ~ \mathbb{I} - \frac{\Delta\varepsilon_m}{2}
\begin{pmatrix}
  \cos2\theta_m  & \sin2\theta_m \\
 \sin2\theta_m  & \cos2\theta_m
\end{pmatrix} \right]
\begin{pmatrix}
  \nu_e \\ \nu_a
\end{pmatrix}   ~,
\end{equation}
where $\varepsilon_m = \varepsilon + \tfrac{V}{2}$. The angle $\theta_m$
is known as the \emph{matter mixing angle} and it is defined through the
relations
\begin{equation}\label{MatterVaccumThetaRelation}
\mathcal{A} \sin2\theta_m =  \sin2\theta ~,~
\mathcal{A} \cos2\theta_m =  \cos2\theta - \tfrac{V}{\Delta\varepsilon} ~.
\end{equation}
Consequently, in the presence of matter interaction, we may express the
neutrino flavour states as the superposition of the energy eigenstates given by
\begin{equation}\label{MatterPMNS}
    \begin{pmatrix}
        \nu_e \\ \nu_a
    \end{pmatrix}=
        \begin{pmatrix}
            \cos\theta_m  & \sin\theta_m \\
            -\sin\theta_m  & \cos\theta_m
        \end{pmatrix}
    \begin{pmatrix}
        \psi_{m1} \\ \psi_{m2}
    \end{pmatrix} ~,
\end{equation}
where $\psi_{m1}$ and $\psi_{m2}$ are the energy eigenstates with the 
eigenvalues $\varepsilon_{m1}$ and $\varepsilon_{m2}$ respectively.

\subsubsection{Adiabatic approximation}

The matter mixing angle $\theta_m$ together with the equation 
(\ref{FlavourEvolutionEqMatter}) is strictly valid for a constant potential 
$V$, requiring constant (matter) electron density $n_e$ as well as a constant 
metric function $\Phi$. However, these equations may still be used for a slowly 
varying density medium if the variation is subject to the adiabatic condition 
\cite{Mikheyev:1985zog, AM}, given by
\begin{equation}\label{AdiabaticCondition}
\left|\frac{\dot{\theta}_m}{\Delta\varepsilon_m} \right| \ll 1 ~.
\end{equation}
If the condition (\ref{AdiabaticCondition}) is satisfied then the transitions 
between the energy eigenstates $\psi_{m1} \longleftrightarrow \psi_{m2}$ can be 
neglected and the energy eigenstates can be considered to be propagating 
independently of each other. It also requires that the matter interaction be 
perturbative in nature \emph{i.e.} $|\varepsilon_{m} - \varepsilon| \ll
\varepsilon$. Consequently, we may approximate the trajectories of neutrinos
inside a matter medium to follow the geodesics, implying the eigenvalues
$\varepsilon_{m1}$ and $\varepsilon_{m2}$ to remain conserved during the
propagation.

\subsubsection{Applicability of matter mixing for a neutron star}

In view of considerable uncertainties in our understanding of the dense nuclear
matter, it is an open issue whether the interaction (\ref{InteractionAction})
considered in the study of solar neutrinos would remain to be the most relevant
interaction even for the neutron stars. If it is so then from the FIG.
\ref{fig:VacuumOscLength}, we note that at the relevant scale for the neutron
star radius, $\Delta\varepsilon \sim 10^{-11}$ eV. It corresponds to the
characteristic density being $n_{\varepsilon} \sim 10^{-13}$ fm$^{-3}$. On the
other hand, inside a neutron star the scale of electron density being $n_e \sim
$ fm$^{-3}$ corresponds to $V \sim 10^2$ eV. Therefore, the parameter
$\mathcal{A}$ inside a typical neutron star becomes $\mathcal{A} \sim 10^{13}$.
Further, if we assume $(\dot{V}/V) \sim (1/R)$ then the adiabatic approximation
ratio becomes $(\dot{\theta}_m/\Delta\varepsilon_m) \sim (\Delta\varepsilon/V^2
R) ~\sim 10^{-25} \ll 1$. It shows that adiabatic condition
(\ref{AdiabaticCondition}) could be satisfied inside a typical neutron star.

\subsubsection{Transition probability of neutrinos}

In the presence of matter interaction, the transition probability of an 
initially electron neutrino state $\nu_{e|0} = \cos\theta_m^0 \psi_1(0) + 
\sin\theta_m^0 \psi_2(0)$ transforming into a non-electron neutrino state 
$\nu_{a|L} = -\sin\theta_m^L \psi_1(L) + \cos\theta_m^L \psi_2(L)$ after 
propagating through a baseline $L$ in a slowly varying matter density medium 
where the adiabatic condition (\ref{AdiabaticCondition}) holds, is given 
by \cite{MSWdev}
\begin{eqnarray}\label{TransitionProbabilityMatter}
P(e\rightarrow a) &=& \sin^2\theta_m^L\cos^2\theta_m^0 +
\cos^2\theta_m^L\sin^2\theta_m^0 \nonumber \\
&&  -~\tfrac{1}{2} \sin 2\theta_m^L \sin 2\theta_m^0 \cos(\Delta\phi_{osc}) ~.
\end{eqnarray}
We note that in a constant density medium where $\theta_m^L = \theta_m^0$
or in the vacuum where $\theta_m = \theta$, the transition probability 
(\ref{TransitionProbabilityMatter}) reduces to the expression 
(\ref{TransitionProbabilityDef}) as expected.

\section{Neutron Stars}

Inside a neutron star under $\beta$-equilibrium, the constituent nucleons go
through the processes of $\beta$-decay and inverse $\beta$-decay. These
processes are known as the Urca processes and they produce neutrinos or
anti-neutrinos of electron flavour only. One such process, known as the direct
Urca process, leads to generation of electron-flavour neutrinos through the
process $n\to p + e^{-} + \bar{\nu}_e$ and $p\to n + e^{+} + \nu_e$. These
neutrinos then propagate out of the neutron stars carrying away energy.
Consequently these neutrino generation processes form a dominant cooling
mechanism for the neutron stars. There are considerable uncertainties in
understanding the efficiencies of different Urca processes. However, each of
these processes primarily leads to the generation of only electron-flavour
neutrinos. This particular aspect would play an important role for using
the neutrino oscillations as a probe for constraining the nuclear matter EOS of
the neutron stars.

\subsection{Flavour composition of emitted neutrinos}

As discussed before, the Urca processes inside a neutron star are known to
produce only electron-flavour of neutrinos. However, a certain fraction of
these electron neutrinos would transform into non-electron neutrinos due to the
flavour oscillations during their propagation. The flavour composition of
emitted neutrinos depends on the phase difference between the corresponding
energy eigenstates (\ref{DeltaPhaseCurvedA1}). Besides, the probability of
direct Urca process occurring near the stellar core is much higher
which would lead to a significantly higher neutrino luminosity from the core in
comparison to the stellar crust \cite{Brown_2018}. In addition, the energy
scales associated with the emitted neutrinos from the crust are also expected to
be different compared to those from the core. Therefore, for simplicity, we
shall assume these
electron-neutrinos are predominantly produced near the core of the star. So the
phase difference near the stellar surface when the neutrinos are propagating out
of the neutron star, can be expressed as
\begin{equation}\label{DeltaPhaseCurvedAtR}
\Delta \phi_{osc} \simeq \frac{\Delta m^2}{2\varepsilon} (R\gamma_m)
+ \Delta\varepsilon (R\gamma_{\varepsilon})  ~,
\end{equation}
where 
\begin{equation}\label{GammaRandE}
\gamma_m = \frac{1}{R} \int_0^R dr~ e^{\Phi + \nu}  ~,~
\gamma_{\varepsilon} = \frac{1}{R} \int_0^R dr~ e^{\nu- \Phi}  ~.
\end{equation}
If we use the approximation $\Delta\varepsilon = (\Delta m^2/2\varepsilon)$ then
the phase difference becomes
\begin{equation}\label{DeltaPhaseCurvedAtRAvg}
\Delta \phi_{osc} =  \frac{\Delta m^2}{\varepsilon} (R\gamma)
~~\textrm{with}~~
\gamma = \frac{1}{2} (\gamma_m  + \gamma_{\varepsilon}) ~.
\end{equation}
In the flat spacetime limit \emph{i.e.} $\Phi\to0$, $\nu\to 0$, the factor
$\gamma\to 1$. However, in a curved spacetime the factor $\gamma$ can be
different from $1$ depending on the interior metric solution of the star.
In summary, the flavour composition of emitted neutrinos from a neutron star
near the stellar surface carry information about the interior spacetime metric. 
This aspect can be used to put constraints on the possible nuclear matter EOS 
of a given neutron star.

\subsection{Extraction of the factor $\gamma$}

\subsubsection{Vacuum or constant electron density medium}

From the equations (\ref{TransitionProbabilityDef},
\ref{DeltaPhaseCurvedAtRAvg}) we note that in principle if we could observe
the flavour composition of emitted neutrinos at two different energy scales
say $\varepsilon^{(1)}$ and $\varepsilon^{(2)}$ then we could eliminate the
mixing angle $\theta$ from the determination of the factor $\gamma$. In
particular, we could express
\begin{equation}\label{ProbabilityRatioCurved}
\frac{P^{(1)}}{P^{(2)}} = \frac{\sin^2(\Delta\phi^{(1)}_{osc}/2)}
{\sin^2(\Delta\phi^{(2)}_{osc}/2)} ~,
\end{equation}
where $P^{(j)} =  P^{(j)}(e\rightarrow a)$ and $\Delta\phi^{(j)}_{osc} =
(\Delta m^2 R\gamma)/\varepsilon^{(j)}$ for $j=1,2$. In other words, for a
given neutron star, one could extract the factor $\gamma$ from the determination
of the ratio of normalized counts of non-electron neutrinos emitted at two
different energy scales near the stellar surface.

\subsubsection{Varying electron density medium}

Similar to the case of vacuum or constant density medium, one could extract the
factor $\gamma$ even for a medium where density variations obey the adiabatic
approximation. However, in this case one would need to observe the flavour
composition of emitted neutrinos for at least \emph{three} different
energy scales say $\varepsilon^{(1)}$, $\varepsilon^{(2)}$, and
$\varepsilon^{(3)}$. Consequently, we could eliminate the mixing angle from the
determination of the factor $\gamma$ as
\begin{equation}\label{ProbabilityRatioCurvedM}
\frac{P^{(1)} - P^{(2)}}{P^{(1)} - P^{(3)}}
= \frac{\cos(\Delta\phi^{(1)}_{osc})/\varepsilon^{(1)} -
\cos(\Delta\phi^{(2)}_{osc})/\varepsilon^{(2)}}
{\cos(\Delta\phi^{(1)}_{osc})/\varepsilon^{(1)} -
\cos(\Delta\phi^{(3)}_{osc})/\varepsilon^{(3)}}  ~,
\end{equation}
where $\Delta\phi^{(j)}_{osc} = (\Delta m^2 R\gamma)/\varepsilon^{(j)}$ for
$j=1,2,3$. Here we have neglected the terms which are 
$\mathcal{O}(\mathcal{A}^{-2})$ for large values of  $\mathcal{A}$. As earlier, 
for a given neutron star, one could extract the factor $\gamma$ from the 
determination of the ratio of the normalized differential counts of non-electron 
neutrinos emitted near the stellar surface by considering at least three 
different energy scales.

\subsection{Constraint on the equation of states}

We have already argued that the flavour composition of emitted neutrinos
near the stellar surface depends, apart from the properties of neutrinos and the 
stellar radius $R$, also on the factor $\gamma$ which in turns depends on the 
interior metric solution. Therefore, the values of the factor $\gamma$ 
estimated from the observations can be contrasted with the values which are 
computed for different possible nuclear matter EOS of a given neutron star.

\subsubsection{Models of equation of states}

In order to elucidate the usage of neutrino oscillations as a probe, in this
section we compute the factor $\gamma$ by considering three different choices of
the nuclear matter EOS for a neutron star. The first EOS that we choose is the
standard \emph{polytropic} EOS of the form $P \propto \rho^{1 + 1/k}$. For
convenience, here we consider a parametric form for the energy density $\rho$ as
$\rho = m_n n$ where $n$ may be viewed as the nucleon number density and $m_n$
being the average mass of the nucleons. Here we choose $m_n$ to be the mass of
the neutrons. Further, we introduce a constant $b = (3\pi^2/m_n^3)$ so that the
term $(bn)$ is a dimensionless quantity. Therefore, we can express the energy
density $\rho$ and the pressure $P$ corresponding to the polytropic EOS in a
parametric form as
\begin{equation}\label{PolytropicEOS}
\rho = \frac{m_n^4}{3\pi^2} (bn) ~~,~~
P = c_k  \frac{m_n^4}{3\pi^2} (bn)^{1 + \tfrac{1}{k}} ~,
\end{equation}
where $k$ and $c_k$ are referred to as the \emph{polytropic index} and the
\emph{polytropic coefficient} respectively.

Additionally, we consider two variants of the nuclear matter EOS described by 
the $\sigma-\omega$ model of nuclear interaction and computed using thermal 
quantum field theory in a curved spacetime \cite{hossain2021higher, 
hossain2022methods}. The first variant of the equation of state is computed in a 
flat spacetime and we refer to it as the $\sigma-\omega$ \emph{flat} EOS. The
second variant of the equation of state is computed by considering the curved 
spacetime of a given neutron star which we refer to as the $\sigma-\omega$ 
\emph{curved} EOS. For both these model EOS, the nuclear matter is composed of 
\emph{neutrons}, \emph{protons} and \emph{electrons} in $\beta$-equilibrium and 
interaction between the baryons are mediated by the $\sigma$ and $\omega$ 
mesons.

\subsubsection{Neutron star with mass $1.4$ $M_{\odot}$}

\begin{figure}
\includegraphics[height=5.8cm, width=8cm]{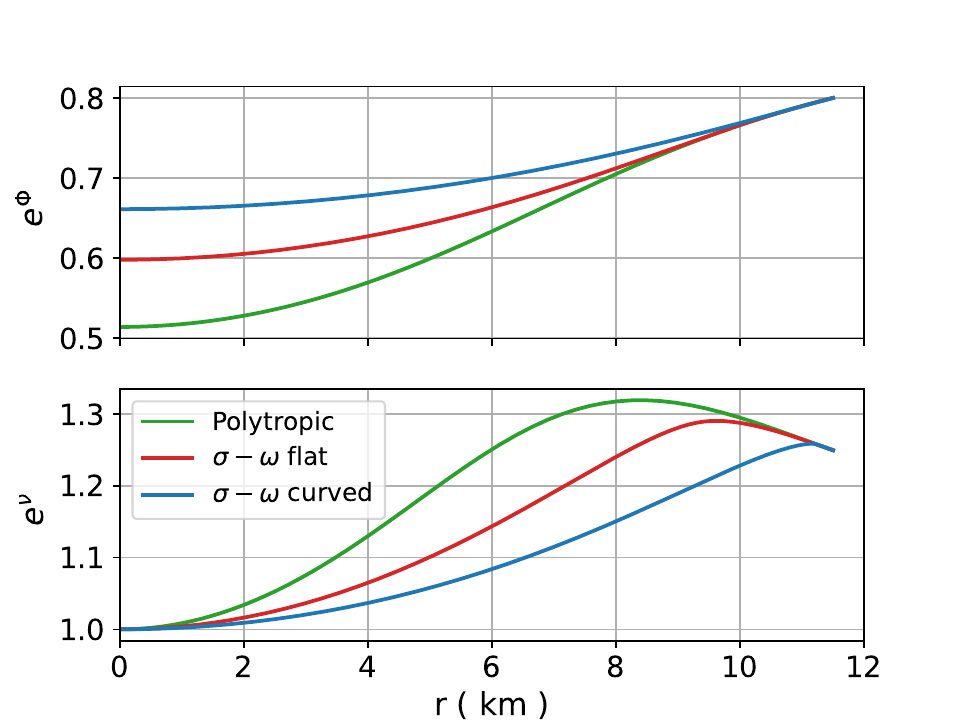}
\caption{The radial variations of the metric functions $e^{\Phi}$ and
$e^{\nu}$ inside a neutron star for three different EOS. The EOS parameters,
Set 1, are chosen to ensure that all three EOS lead to a neutron star having
mass $M=1.40$ $M_{\odot}$ and radius $R=11.50$.}
\label{fig:PhiNuRadial14}
\end{figure}

\begin{figure}
\includegraphics[height=5.8cm, width=8cm]{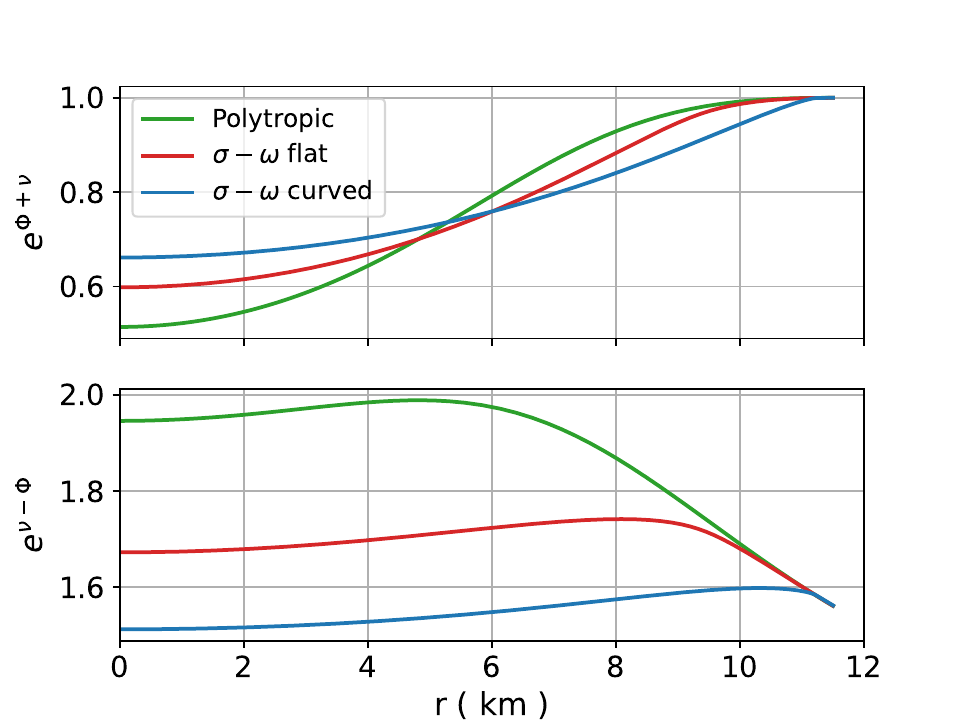}
\caption{The radial variations of the integrand $e^{\Phi+\nu}$ and
$e^{\nu-\Phi}$ inside a neutron star having mass $M=1.40$ $M_{\odot}$ and
radius $R=11.50$ for three different EOS as given in Set 1.}
\label{fig:EPhiENuRadial14}
\end{figure}

Let us consider a neutron star having mass $M=1.40$ $M_{\odot}$ and radius
$R=11.50$ km. Such a neutron star can be obtained from multiple EOS with
different values of their parameters. A set of parameters that lead to the given
neutron star due to the polytropic EOS, the $\sigma-\omega$ flat EOS and the
$\sigma-\omega$ curved EOS are tabulated as `Set 1' in the TABLE
\ref{table:Parameters14}.
\begin{table}
\caption{Three different EOS and the associated parameter choices that lead to a
neutron star having mass $M=1.40$ $M_{\odot}$ and radius $R=11.50$ km. For a
given EOS, the parameter choices are not unique. Here the parameter $n_c$ refers
to the central nucleon density. Additionally, we note that for the
$\sigma-\omega$ flat EOS, the nucleon density $n_c=0.684$ fm$^{-3}$, in Set 1,
includes contributions from \emph{neutron} number density $n_{nc}=0.624$
fm$^{-3}$ and \emph{proton} number density $n_{pc}=0.060$ fm$^{-3}$. }
\begin{tabular}{l c c l}
\hline
EOS~~ & Set & $n_c$ & ~~Parameters \\
 &  & (fm$^{-3}$) &  \\ \hline
Polytropic & 1 & 1.652 & $k=1.45, ~~~c_k=0.342$ \\
$\sigma-\omega$ flat  & 1 & 0.684 & $\gtomega=14.89, \zeta=0.046,
~\gtsigma=14.94$ \\
$\sigma-\omega$ curved & 1 & 0.615 & $\gtomega=10.90, \zeta=0.023,
~\gtsigma=13.10$ \\
\hline
Polytropic & 2 & 1.707 & $k=1.45, ~~c_k=0.345$ \\
$\sigma-\omega$ flat  & 2 & 0.666 & $\gtomega=14.87, \zeta=0.045,
~\gtsigma=14.97$ \\
$\sigma-\omega$ curved & 2 & 0.613 & $\gtomega=10.89,\zeta=0.023,
~\gtsigma=13.09$ \\
\hline
\label{table:Parameters14}
\end{tabular}
\end{table}
The resultant mass, radius and the factor $\gamma$ are tabulated in the TABLE 
\ref{table:MassRadiusGamma14}.
\begin{table}
\caption{The factor $\gamma$ for different EOS choices as given in the TABLE
\ref{table:Parameters14}.}
\begin{tabular}{l c c c c c c cc}
\hline
EOS & Set & ~~$M$~~ & ~~$R$~~ & ~~~~$\gamma_{m}$~~~~ &
~~~~$\gamma_{\varepsilon}$~~~~ & ~~~~$\gamma$~~ \\
 &  & $(M_{\odot})$ & $(km)$~ &  & & \\ \hline
Polytropic             & 1 & 1.40 & 11.50 & 0.767 & 1.885 & 1.326 \\
$\sigma-\omega$ flat   & 1 & 1.40 & 11.50 & 0.776 & 1.695 & 1.236 \\
$\sigma-\omega$ curved & 1 & 1.40 & 11.50 & 0.783 & 1.550 & 1.167 \\
\hline
Polytropic             & 2 & 1.41 & 11.42 & 0.763 & 1.911 & 1.337 \\
$\sigma-\omega$ flat   & 2 & 1.39 & 11.55 & 0.779 & 1.679 & 1.229 \\
$\sigma-\omega$ curved & 2 & 1.39 & 11.48 & 0.785 & 1.546 & 1.165 \\
\hline
\label{table:MassRadiusGamma14}
\end{tabular}
\end{table}
Although all 3 different EOS lead to a neutron star having the same
mass $M$ and radius $R$ hence the same compactness but the resultant $\gamma$ 
factor differs from each other due to having different interior metric 
solutions. The radial variations of the metric solutions, for the Set 1 
parameters, are plotted in the FIG. \ref{fig:PhiNuRadial14} whereas the radial 
variations of integrand $e^{\Phi+\nu}$ and $e^{\nu-\Phi}$ are plotted in the 
FIG. \ref{fig:EPhiENuRadial14}.

As an illustration, we note that if neutrinos are observed at two different
energies say $\varepsilon^{(1)} = 4$ MeV and $\varepsilon^{(2)} = 8$ MeV then
for $\Delta m^2 = 8 \times 10^{-5}$ eV$^2$ along with the parameters given in
the Set 1 of the TABLE \ref{table:MassRadiusGamma14}, the probability ratio
(\ref{ProbabilityRatioCurved}) takes the values $3.43$, $3.50$ and $3.56$ for
the polytropic, the $\sigma-\omega$ flat, and the $\sigma-\omega$ curved EOS
respectively. In principle, such a ratio could be compared with the
observations. Additionally, from the TABLE \ref{table:MassRadiusGamma14}, we
note that for a specific EOS, a small variation of the mass and the radius lead
only to a small variation of the factor $\gamma$. We also note that if one
considers the neutrino luminosity of a neutron star core to be $10^{31}$ J
s$^{-1}$ as estimated in \cite{Brown_2018}, then the flux of non-electron
neutrinos at the stellar surface, implied by the equations
(\ref{TransitionProbabilityDef}, \ref{DeltaPhaseCurvedAtRAvg}) for $\varepsilon
= 4$ MeV, $\Delta m^2 = 8 \times 10^{-5}$ eV$^2$, $\sin^2 2\theta = 0.83$
\cite{Smirnov2003}, and with the Set 1 parameters in the TABLES
\ref{table:Parameters14} and \ref{table:MassRadiusGamma14}, would be
$3.8 \times 10^{33}$ m$^{-2}$ s$^{-1}$,
$3.4 \times 10^{33}$ m$^{-2}$ s$^{-1}$, and
$3.1 \times 10^{33}$ m$^{-2}$ s$^{-1}$ for the polytropic,
$\sigma-\omega$ flat and $\sigma-\omega$ curved EOS respectively.

In addition, we note that for a neutron star described by the Set 1 parameters
given in the TABLE \ref{table:Parameters14} together with $\Delta\varepsilon =
4.0\times10^{-11}$ eV and assuming the electron density to be one-tenth of
nucleon density, the adiabatic approximation ratio
$(\dot{\theta}_m/\Delta\varepsilon_m)$ can be estimated to be $1.6 \times
10^{-24}$, $9.1 \times 10^{-24}$, and $1.1 \times 10^{-23}$ for the
polytropic, $\sigma-\omega$ flat and $\sigma-\omega$ curved EOS respectively. It
shows that one can consistently use the adiabatic approximation
(\ref{AdiabaticCondition}) for the model EOS employed here.

\subsubsection{Neutron star with mass $2.0$ $M_{\odot}$}

Let us consider another example of a neutron star having mass $M=2.00$
$M_{\odot}$ and radius $R=12.50$ km.  A chosen set of parameters that lead to
the given neutron star due to the polytropic EOS, the $\sigma-\omega$ flat EOS
and the $\sigma-\omega$ curved EOS are tabulated in the TABLE
\ref{table:Parameters20}.
\begin{table}
\caption{Three different EOS and the associated parameters that lead to a
neutron star having mass $M=2.00$ $M_{\odot}$ and radius $R=12.50$ km. As
earlier, here too the parameter sets are not unique.}
\begin{tabular}{l c l}
\hline
EOS ~~~~~~~~~ & $n_c$(fm$^{-3}$) & ~~Parameter set \\
\hline
Polytropic            & 1.319  & $k=1.16, ~~c_k=0.790$ \\
$\sigma-\omega$ flat   & 0.640 & $\gtomega=14.80, \zeta=0.014,
~\gtsigma=15.73$ \\
$\sigma-\omega$ curved & 1.244 & $\gtomega=14.66, \zeta=0.068,
~\gtsigma=14.88$ \\
\hline
\label{table:Parameters20}
\end{tabular}
\end{table}
The resultant mass, radius and the factor $\gamma$ is tabulated in the TABLE
\ref{table:MassRadiusGamma20}.
\begin{table}
\caption{The factor $\gamma$ corresponding to the EOS choices given in the TABLE
\ref{table:Parameters20}.}
\begin{tabular}{l c c  c c c cc}
\hline
EOS & ~~$M$~~ & ~~$R$~~ & ~~~~$\gamma_{m}$~~~~ &
~~~~$\gamma_{\varepsilon}$~~~~ & ~~~~$\gamma$~~ \\
 & $(M_{\odot})$ & (km) ~ &  & & \\ \hline
Polytropic             & 2.00 & 12.50 & 0.669 & 2.416 & 1.542 \\
$\sigma-\omega$ flat   & 2.00 & 12.50 & 0.685 & 2.073 & 1.379 \\
$\sigma-\omega$ curved & 2.00 & 12.50 & 0.685 & 2.071 & 1.378 \\
\hline
\label{table:MassRadiusGamma20}
\end{tabular}
\end{table}

\begin{figure}
\includegraphics[height=5.9cm, width=8cm]{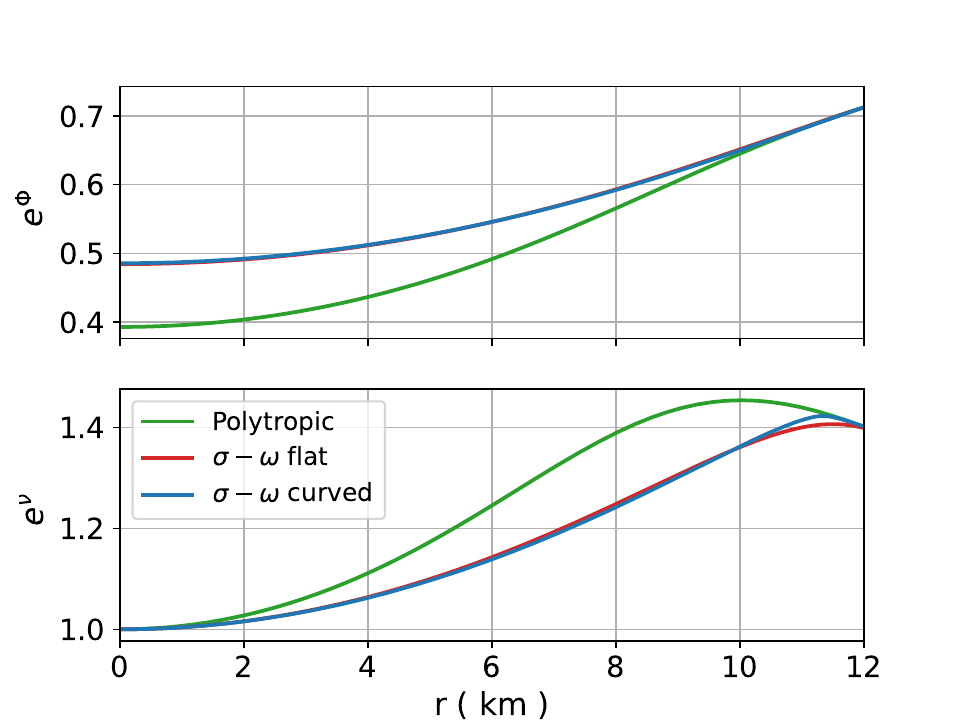}
\caption{The radial variations of the metric functions $e^{\Phi}$ and $e^{\nu}$
inside a neutron star having mass $M=2.00$ $M_{\odot}$ and radius $R=12.50$ km
for three EOS choices given in the TABLE \ref{table:Parameters20}. We note here
that the metric functions nearly overlap for the $\sigma-\omega$ flat and curved
EOS.}
\label{fig:PhiNuRadial20}
\end{figure}

Unlike the previous example, here we note that despite having different
parameter sets, the $\sigma-\omega$ flat EOS and the $\sigma-\omega$ curved EOS
cannot be well distinguished from comparison of the factor $\gamma$. However,
one can still differentiate between the polytropic EOS and the $\sigma-\omega$
EOS. This aspect can also be seen from the radial variation of the metric
functions in the FIG. \ref{fig:PhiNuRadial20}. This example also shows the
limitation of using neutrino oscillations for differentiating EOS of neutron
stars. Clearly, there are regions in the parameter space where the method
described here may not be able to distinguish different possible nuclear matter
EOS.

\section{Discussions}

In summary, we have shown that the phenomena of neutrino oscillations and the 
flavour mixing could be used as a probe for differentiating possible equation of 
states of nuclear matter of a neutron star if one could determine the flavour
composition of emitted neutrinos near the stellar surface. Such possible usage 
of neutrino oscillations stems from the fact that the transition probabilities
among different neutrino flavours are dependent on the interior stellar metric
through which these neutrinos propagate out from the stellar interior. On the
other hand, different equation of states of nuclear matter implies different
interior metric solutions even for a neutron star having the same mass and
radius. The method described here could work even for an isolated neutron star
where constraining its nuclear matter equation of states through the tidal
deformation method is not possible.

Nevertheless, it is acknowledged that the direct determination of the flavour
composition of emitted neutrinos near the stellar surface would be a challenging
task for a distant observer and one may have to look for possible indirect
methods. Observational data indicates that the neutron stars have a thin
layer of atmosphere \cite{2009Natur.462...71H, zavlin2002}. An indirect method
for determination of the flavour composition could be to look for possible
signatures emanating from the interactions of non-electron neutrinos with the
stellar atmosphere.

In our analysis, we have assumed that the neutrinos are described by the Dirac
spinor. However, there exists a debate in the literature whether the neutrinos
should be considered as Majorana fermions. Besides, we have considered only the
left-handed neutrinos to undergo interactions with the background electron
density.

\begin{acknowledgments}
SB thanks IISER Kolkata for supporting this work through a masters fellowship.
GMH acknowledges support from the grant no. MTR/2021/000209 of the SERB, 
Government of India.
\end{acknowledgments}

\bibliographystyle{apsrev}

\end{document}